\documentclass{iaus}
\usepackage{graphicx}

\title[Quantitative spectroscopy of close binary stars] 
{Quantitative spectroscopy \\ of close binary stars}

\author[K.~Pavlovski \& J.~Southworth]   
{K.~Pavlovski$^{1,2}$ \&  J.~Southworth$^2$}

\affiliation{
$^1$Department of Physics, Faculty of Science, University of Zagreb, Croatia \\[\affilskip]
$^2$Astrophysics Group, Keele University, Staffordshire, ST5 5BG, UK
}

\pubyear{2011}
\volume{282}  
\pagerange{119--126}
\jname{From Interacting Binaries to Exoplanets: Essential Modelling Tools}
\editors{A.C. Editor, B.D. Editor \& C.E. Editor, eds.}
\begin{document}

\maketitle

\begin{abstract}
The method of spectral disentangling has now created the opportunity for studying the chemical
composition in previously inaccessible components of binary and multiple stars. This in turn makes
it possible to trace their chemical evolution, a vital aspect in understanding the evolution
 of stellar systems. We review different ways to reconstruct individual spectra from eclipsing
and non-eclipsing systems, and then concentrate on some recent applications to detached
binaries with high-mass  and intermediate-mass stars, and Algol-type mass-transfer systems.
\end{abstract}

\section{Introduction}

The structure and evolution of a star is determined by its mass and chemical composition. Hence,
the position of the star in the HR diagram is uniquely defined only when its mass and bulk
metallicity are known. Eclipsing and double-lined spectroscopic binaries in a detached
configuration remain the primary source of directly measured fundamental stellar quantities:
mass and radius. Modern observational techniques are currently able to reach precisions of
1--2\% in these parameters, a prerequisite for testing theoretical stellar evolutionary models.
In their critical survey and compilation of available results, Torres, Andersen \& Gimenez (2010)
 selected 95 detached eclipsing binaries (dEBs) which satisfied a reasonable criterion of 3\%
in precision for both quantities. The effective temperatures of the component stars cannot be
determined directly from the analysis of light or radial velocity curves, making $T_{\rm eff}$
a less well-determined quantity.

The metallicity has been determined for the components of only about 21 of the 95 binaries in
the sample of Torres et al.\ (2010). Usually, metallicity is derived from fitting the overall
spectrum of the stellar system. Detailed abundance determinations have been accomplished for
only {\it four} binaries: directly from the spectrum of $\beta$ Aur (Lyubimkov et 
al.\ 1996) and CV\,Vel (Yakut et al.\ 2007), and
from disentangled component spectra of V578\,Mon (Pavlovski \& Hensberge 2005) and V453\,Cyg
(Pavlovski \& Southworth 2009).

Tremendeous advances in observational stellar spectroscopy, both in spectral resolution, quantum
efficiency and detector linearity, allow new opportunities in binary star research. Reconstructing
 the individual spectra of the components opens a new window on these stars: detailed spectral
analysis, atmospheric diagnostics and determination of atmospheric chemical compositions
(Pavlovski 2004, Pavlovski \& Hensberge 2005). This enables a fine probing of stellar evolutionary
models and proper calibrations of fundamental stellar quantities (Pavlovski \& Southworth 2009,
Pavlovski et al.\ 2009, 2011). The advantages of analysing disentangled spectra are manifold.

\section{Renormalisation of disentangled component spectra}

More than a dozen methods have been invented to obtain separate spectra of the components of
multiple systems from a series of composite spectra obtained at a range of orbital phase.
Pavlovski \& Hensberge (2010) divided them into three basic categories: (i) spectral separation;
(ii) spectral disentangling; and (iii) spectroastrometric splitting. In all cases the isolation
of individual spectra exploits the variable Doppler shift of the stars.

In {\em spectral separation} techniques radial velocities (RVs) are input data obtained from
another source (e.g.\ cross-correlation). The method most widely used in this category is
Doppler tomography, developed by Bagnuolo \& Gies (1991). The technique of {\em spectral
disentangling} ({\sc spd}; Simon \& Sturm 1994, Hadrava 1995) refers to the situation where
both the individual spectra and the orbital parameters of the components are measured,
simultaneously and self-consistently. The {\em spectroastrometric splitting} method accesses
the spatial information present in long-slit spectra of partially resolved components
(see Bailey 1998; Wheelwright, Oudmaijer \& Schnerr 2009, and references therein).

The principles of these methods are outlined in the review papers by Hadrava (2004, 2009),
Hensberge \& Pavlovski (2007), and Pavlovski \& Hensberge (2010). Without going into details
we record here some new
developments by Konacki et al.\ (2010), Folsom et al.\ (2010) and Kolbas \& Pavlovski (this Volume).

Spectral disentangling or separation techniques return individual spectra which are either in
the common continuum of the system, or on an arbitrary (generic) continuum. The reason for this
is a basic principle of {\sc spd}: there is no information on the continuum light ratio of the
stars if disentangling is performed on spectra where the light ratio is constant.
The absolute spectral line depths are then unknown because the continuum level is unknown.
The solution is to obtain spectra during eclipse, as information on the continuum level is
then available. Spectra during eclipse can be difficult to secure for reasons including the
scheduling of observations and the lower brightness of the system at eclipse times. However,
the main difficulty lies in the Rossiter-McLaughlin effect. This distorts the line profiles
and thus violates a basic assumption of spectral disentangling. The Rossiter-McLaughlin can
be only avoided by studying totally-eclipsing systems. In practice we usually have a situation
in which the continuum light ratios do not vary between spectra, and {\sc spd} has to be
performed with arbitrary or generic light dilution factors. Renormalisation then requires
light factors from other sources.

It is expected that the most reliable light dilution factors come from analysis of the light
curves of dEBs. This is true, but it depends on particular cases, as discussed below. When
eclipses do not occur, or observational data do not cover them, we must rely on information
contained in the disentangled spectra themselves. Here we list several different approaches
 recorded in the literature:

\begin{itemize}

\item For eclipsing binaries, light factors may be available from the time-independent
dilution of spectral lines or from light curves (e.g.\ Hensberge, Pavlovski \& Verschueren 2000).
The quality of the photometry, the configuration and geometry of the binary system, or the
presence of a third star, could make the light factors uncertain.

\item The light factors can be determined from the time-dependent dilution of the spectral
lines as additional free parameters in {\sc spd} calculations ({\em line photometry}; Hadrava 1997).
Any intrinsic line profile changes (pulsations, spots, Rossiter-McLaughlin effect, etc.)
violate a basic assumption of spectral disentangling, and can cause erroneous results.

\item If the system is not eclipsing, some {\em physical considerations} can be used to
renormalise individual disentangled spectra. In the study of the non-eclipsing triple system
DG\,Leo, Fr\'{e}mat et al.\ (2005) successfully used the very deep Ca~{\sc ii}\,K line profiles.
The requirement that the core should not dip below zero flux for any of the component spectra
(in this particular case all components are of similar spectral type) imposed very strong
constraints on the light factors. Light dilution factors can also be estimated from line
depths, or equivalent width ratios, once the atmospheric characteristics are fixed
(c.f.~Mahy et al.\ 2011). Use of line depth or line intensity
ratios require chemical abundances to be specified {\it a priori}, making this option
somewhat uncertain.

\item Separated or generic disentangled spectra contain information on the intrinsic spectra.
Just as in the way in which information on effective temperature ($T_{\rm eff}$) and surface
 gravity ($\log g$) are extracted from renormalised spectra, it is possible to recover also
light factors by constrained multi-parameter line-profile fitting of both disentangled spectra.
Tamajo, Pavlovski \& Southworth (2011) have implemented this idea in the code {\sc genfitt}.
Simulations using synthetic spectra showed that for a reasonable signal to noise ratio of S/N
$\ge$ 100 {\em constrained genetic fitting} returns reliable $T_{\rm eff}$ and $\log g$ for
each star, plus their light factor. It is so called because the optimal fitting is constrained
by requiring the sum of the light factors to be unity. $\log g$ can be derived for the stars
in close binaries with precisions of 0.01\,dex, a crucial advantage in breaking the degeneracy
between $T_{\rm eff}$ and $\log g$ for Balmer line profiles. Comparison of the light factors
derived by constrained genetic fitting and from light curve analyses is given by Pavlovski
et al.\ (2009), Tamajo et al.\ (2011), and Southworth et al.\ (2011b) for some real-world cases.

\item The procedure described in the previous item could be generalised in the way that part
or whole of the disentangled spectrum is fitted by theoretical spectra. Tkachenko, Lehmann \&
Mkrtichian (2009, 2010) have used this idea and fitted a large portion of spectra by gridding
with precomputed theoretical spectra for a large range of $T_{\rm eff}$, $\log g$ and
metallicity. A similar technique has been applied by Torres et al.\ (2011) but with restriction
to solar abundances, only.

\end{itemize}

\section{Chemical composition from reconstructed spectra}

To determine the chemical composition of a stellar atmosphere, it is first neccessary to
specify an appropriate model atmosphere. The model is described by $T_{\rm eff}$, $\log g$,
and metallicity ([M/H]). For most eclipsing binaries $\log g$ can be determined, from analysis
 of light and velocity curves, with a precision and accuracy up to an order of magnitude larger
than for single stars. This considerably facilitates the determination of $T_{\rm eff}$,
and side-steps the degeneracy between $T_{\rm eff}$ and $\log g$ for hot stars. Moreover,
a reliable estimate of $T_{\rm eff}$ is possible directly from {\em constrained fitting}
when the light factors are not well determined from external sources (Pavlovski et al.\ 2009,
Tamajo et al.\ 2011). This also makes it possible to estimate $T_{\rm eff}$ and $\log g$
for non-eclipsing binaries, and stipulate additional constraints in complementary solutions
with interferometric observations. The first estimate of $T_{\rm eff}$ can be used in an
iterative cycle for fine-tuning light curve and {\sc spd} solutions, as was described by
Hensberge et al.\ (2000) and further elaborated by Clausen et al.\ (2008).

Detailed abundance work makes possible further improvements in $T_{\rm eff}$ determination
through ionisation balance. We have successfully employed the Si\,{\sc ii}\,/\,Si\,{\sc iii}
ionisation balance in high-mass binaries (Pavlovski \& Southworth 2009), and
the Fe\,{\sc i}\,/\,Fe\,{\sc ii} balance in intermediate-mass binaries (work in preparation).
Such an improvement in determination of $T_{\rm eff}$ is an important ingredient if one
wants to use binaries for distance determination, in particular for the galaxies in
Local Group (e.g.\ Harries et al.\ 2003; North et al.\ 2011).

\subsection{Detached binary stars}

{\bf High-mass stars}. \
Despite considerable theoretical and observational efforts, some important pieces of
the jigsaw of stellar structure and evolution remain unclear or missing. Meynet \& Maeder
(2000) and Heger \& Langer (2000) found that rotationally-induced mixing and magnetic fields
could cause substantial changes in theoretical predictions. Some of these concern evolutionary
changes in the chemical composition of stellar atmopsheres. In close binaries, tidal effects
further complicate this picture (De Mink et al.\ 2009), and pose a big challenge for
observational confirmation.

Our observational project on the chemical evolution of high-mass stars in close binaries
is directed toward tracing predicted changes in the photospheric abundance pattern due
to rotational mixing. In Pavlovski et al.\ (in preparation) we summarise our results
for fourteen high-mass stars in eight dEBs, plus some additional high-mass stars in
binaries studied from disentanled spectra (Simon et al.\ 1994, Sturm \& Simon 1994,
Southworth \& Clausen 2007). Of these, V380\,Cyg (Pavlovski et al.\ 2009),
V621\,Per (Southworth et al.\ 2004), and V453\,Cyg (Pavlovski \& Southworth 2009)
are the most informative as their primary components are evolved either close to or beyond
the terminal-age main sequence. No abundance changes relative to unevolved MS stars
of the same mass have been detected for these components, probably due to their relatively
long orbital periods (De Mink et al.\ 2009).
The study of HD\,48099, an O5.5\,V((f)) + 09\,V binary
system, by Mahy et al.\ (2010) reveals a nitrogen enhancement in the primary star,
but a solar abundance for the secondary. The estimated masses are 55 and 19
M$_\odot$ for the primary and secondary component, respectively. Determination of chemical
composition from disentangled spectra is an important way to constrain theoretical models.

{\bf Intermediate-mass and solar type stars}. \
We have recently constructed detailed abundance studies of late-B and A-type stars in
the close dEBs AS\,Cam and YZ Cas (work in preparation). Abundances are also available
for the $\delta$\,Scuti pulsating components in the binaries DG\,Leo (Fr\'{e}mat et al.\ 2005)
 and HD\,61199 (Harater et al.\ 2008).

Systematic research in FGK stars in binaries, concerning also their chemical composition
and an empirial evaluation of their metallicity, has been initialised by
the late Jens Viggo Clausen and collaborators. Comprehensive study of three F-type
binaries has shown the full power of testing and comparing recent stellar evolutionary
models using eclipsing binaries, provided their abundances are known (Clausen et al.\ 2008).
The same methodology was extended to the solar-type binary systems V636\,Cen (Clausen et al.\ 2009)
 and NGC\,6791 V20 (Grundahl et al.\ 2008; Brogaard et al., 2011).

\subsection{Algol systems}

One of the many consequences of the first and rapid phase of mass transfer in close binary
systems, and the eventual mass reversal and formation of Algols, is the changes in chemical
composition of the stars involved. In fact, Algols offer an unique opportunity to probe into
stellar interiors since detailed abundance studies of the layers which were once deep inside
the star can give important information on the thermonuclear and mixing processes taking place
during core hydrogen burning (Sarna \& De Greve 1996). Carbon should be depleted in the CNO cycle
 even during a star's MS lifetime, and observational studies have aimed at testing these predictions
(c.f.\ Tomkin 1981 and references therein).

We have started a new observational programme with the aim of deriving detailed abundances from
high-resolution and high-S/N \'echelle spectra using the {\sc spd} technique. We intend to
substantially extend both the number of the elements studied, as number of lines for each element.

Algol ($\beta$\,Per) is the prototype of the class of binary systems in a semidetached configuration,
where the initially more massive and more evolved component fills its inner Roche lobe and transfers
 material to its now more-massive companion. Algol is one of the most frequently studied objects
in the sky, and has been observed at wavelengths ranging from X-rays to radio (c.f.\ Richards et
 al.\ 1988). However, due to difficulties in ground-based observations of such a bright object,
and a lack of modern high-resolution spectroscopy, its stellar and orbital parameters were somewhat hazy.

Since 2007 we have secured 140 high-S/N \'echelle spectra using the FIES spectrograph at the Nordic
Optical Telescope and BOES at Bohyunsan Optical Astronomy Observatory in Korea. The available light
curves are not on their own sufficient to allow the precise quantification of the contribution
of the third component to the total light, which is needed for proper reconstruction of the
disentangled spectra of the components. Therefore, we rely only on spectroscopic information.
Abundances are derived for 15 elements, and are generally close to solar (Kolbas et al., this volume).
We are currently undertaking non-LTE calculations for helium and the CNO elements. We corroborate
the weakness of the Ca\,{\sc ii} lines in the spectrum of the third component, and a slight
underabundance of scandium, both classical indicators of a metallic-lined star. We will be
performing a detailed abundance of our disentangled spectrum for this candidate Am star.

How important an abundance study for understanding stellar evolution in binary systems
is nicely shown by Mahy et al.\ (2011) in a study of the semidetached system LZ\,Cep, an O9\,III + ON9.7\,V
binary. They have found the secondary component, now the less massive star in the system, to be
chemically more evolved than the primary, which barely shows any sign of CNO processing.
Also, considerable changes in the chemical composition which corroborate predictions have been
found for the
components of Plaskett's star (54 + 56 M$_\odot$) by Linder et al.\ (2008). Plaskett's star
is in a post-case-A Roche lobe overflow stage.

{\bf Cool Algols}.
The chemical composition of the primaries of two oEA stars, TW\,Dra and RZ\,Cas, have been
derived by Tkachenko, Lehmann \& Mkrtichian (2009, 2010). This subclass of Algols is known for
cooler A-type primaries which are pulsating with $\delta$\,Scuti characteristics
(Mkrtichian et al.\ 2002). Their analyses of disentangled spectra have shown that these
stars are a normal A-stars with a chemical composition close to solar. Understanding the
chemical composition of the pulsating components in close binaries is an important condition
for proper asteroseismologic diagnostics. In combination with precise stellar parameters derived
from complementary photometric and spectroscopic observations (c.f.~Southwort et al.\ 2011a),
this is a most powerful way for
probing modern models of stellar structure and evolution.
\\

\noindent {\bf Acknowledgements.} KP acknowledges receipt of a Leverhulme Visiting Professorship
which enabled him to work at Keele University, UK, where a part of this work was performed.
JS acknowledges support from the STFC in the form of an Advanced Fellowship. This research is
supported by a grant to KP from the Croatian Ministry of Science and Education.

\end{document}